\documentclass[12pt]{iopart}
\usepackage{graphicx}% Include figure files

\begin{document}

\title{Magnetic phase diagram of the layered superconductor Bi$_{2+x}$Sr$_{2-x}$CuO$_{6+\delta}$ (Bi2201) with  $T_c\approx7$ K}

\author{ S. Salem-Sugui, Jr.$^1$, A.D. Alvarenga$^2$, J. Mosqueira $^3$, J.D. Dancausa $^3$, C. Salazar Mejia $^1$, E. Sinnecker $^1$, Huiqian Luo $^4$, Hai-Hu Wen $^5$}

\address{$^1$Instituto de Fisica, Universidade Federal do Rio de Janeiro,
21941-972 Rio de Janeiro, RJ, Brazil}
\address{$^2$Instituto Nacional de Metrologia Qualidade e Tecnologia, 25250-020 Duque de Caxias, RJ, Brazil}
\address{$^3$LBTS,Universidade de Santiago de Compostela, E-15782, Spain}
\address{$^4$Beijing National Laboratory for Condensed Matter Physics, Institute of Physics, Chinese Academy of Sciences, Beijing  100190,  P. R. China}
\address{$^5$Center for Superconducting Physics and Materials, National Lab of Solid State Microstructures and Department of Physics, Nanjing University, Nanjing 210093, P. R. China}

\begin{abstract}
We report on magnetization measurements performed on a  single crystal of Bi$_{2+x}$Sr$_{2-x}$CuO$_{6+\delta}$ with $T_c\approx7$ K for $H\|c$-axis. The isofield $M(T)$ curves show a large reversible region, with a pronounced rounding effect as $M$ approaches zero which prevents the determination of $T_c(H)$. Deviations from the linear behavior of magnetisation near $T_c(H)$ are studied through the asymptotic behavior of $\sqrt{-M(T)}\propto(T-T_a)^m$, where $m$ is an exponent and $T_a$ an apparent temperature transition. Values of $m$ deviates from the expected mean field value, 1/2, suggesting the importance of phase fluctuations. Resulting values of $T_a$, interpreted as the onset of phase correlations, decrease as field increases showing an upward curvature. Values of $T_c(H)$ are obtained through a two-dimensional critical scaling analysis obeyed by many $M(T)$ curves. The resulting phase diagram do not show upward curvature and lies below the $T_a(H)$ line. The value of $H_{c2}(0)$ estimated from the initial slope d$H_{c2}$/dT is twice the value suggested by the phase diagram. Amplitude fluctuations above $T_a(H)$ are explained in terms of a Ginzburg-Landau approach extended to high reduced temperatures and magnetic fields by the introduction of a total-energy cutoff in the fluctuation spectrum. 
\end{abstract}
\maketitle

\section{Introduction}
The monolayer Bi$_{2+x}$Sr$_{2-x}$CuO$_{6+\delta}$ system, Bi2201, has $T_c\leq10$ K depending on the doping \cite{1,luo}. The system possesses layered structure and Cu-O bonds similar to other cuprates superconductors with much higher values of $T_c$ and upper critical fields at zero temperature, $H_{c2}(0)$. A motivation of the present study, lies on the expected similarity between the phase diagram of Bi2201 \cite{wen} and their parents high-$T_c$ superconductors, as Bi2212 (which displays the pseudo-gap \cite{timusk,gomes1,gomes2}), for which one can only study the linear part (near $T_c$) of the phase diagram with magnetic fields commonly available in laboratories. Another motivation, is the apparent lacking in the literature of fluctuations magnetisation studies in Bi2201, which is in contrast to extensive magnetisation studies performed in Bi2212 \cite{timusk}. Here, we have performed magnetization measurements as a function of magnetic field and temperature in a high-quality single crystal\cite{luo} of Bi$_{2+x}$Sr$_{2-x}$CuO$_{6+\delta}$, with $x=0.1$ and $T_c \approx 7$ K for $H\|c$-axis. Among the studies presented in the literature for Bi2201, it is worth to cite the works of Refs. \cite{chen} and \cite{vedeneev} which obtained magnetic phase diagrams trough resistivity measurements, and the work of Ref. \cite{french} which studied fluctuation conductivity. Results of these works for the optimal doping sample with $x=0.1$ and $T_c \approx 7$~K,  produced quite different values for $H_{c2}(0)$, and for $dH_{c2}/dT$ as well. For instance, values of $H_{c2}(0) \approx 60$~kOe and $dH_{c2}/dT =-16$~kOe/K were found in Ref. \cite{chen} for $H\|c$-axis, while values about 3 times larger were found in Ref. \cite{vedeneev} for the same quantities on a similar sample with same field orientation. Regarding the anisotropy, a $\gamma \approx 3$ were found in Refs. \cite{chen} and \cite{vedeneev} while a much higher value $\gamma\approx 22$ was found in Ref. \cite{french}. The common point of all these works is the existence of an upward curvature in the $H_{c2}(T)~vs.T$ phase diagram. This upward curvature in the phase diagram, which might  be related to the criterion used to extract the values of $H_{c2}(T)$ or $T_c(H)$ from resistivity curves, are claimed to be possibly related to an intrinsic granularity of the Bi2201 system, as observed in overdoped Tl-2201 system.\cite{russian} Moreover, the upward curvature of the critical field in the phase diagram, have been also observed in other high-$T_c$ systems \cite{french}, as for instance in La-Bi2201 with the critical field being extracted from specific heat measurements \cite{wen}. Discussions regarding the critical field, as extracted from resisitivity curves, to be related with the onset of phase coherence \cite{french} point to the need of further studies in these systems. 

In this work, we address the above issue, by performing precise magnetisation measurements in a Bi2201 single crystal. The magnetic phase diagram of the Bi2201 studied sample is obtained through a critical scaling analysis of many isofield magnetisation curves.  The resulting mean field phase diagram do not show any upward curvature, but an analysis of $M(T)$ curves near the transition, considering deviations from the linear behavior of magnetisation with temperature due to phase fluctuations, produced values of an apparent temperature transition, $T_a(H)$ which represents the onset of phase correlations. Values of $T_a(H)$ lies above $T_c(H)$, but below $T_c$, and show an upward curvature and a much higher value of the apparent $dH_{c2}/dT$. The resulting mean field phase diagram shows a $dH_{c2}/dT =-12$~kOe/K, and a  $H_{c2}(0) \approx 30$~kOe, which reasonable agree with values obtained in Ref. \cite{chen} for a sample with similar $T_c$ and same field orientation. The value of  $H_{c2}(0)\approx 60$~kOe, estimated from the WHH formula\cite{whh}, $H_{c2}(0)=-0.69T_c(dH_{c2}/dT)$, is twice the value suggested by the phase diagram. We also observe that a 30~kOe field suppresses any signal of superconductivity above 3~K (the lowest temperature in this work), allowing to obtain the temperature dependence of the normal state magnetisation. 
\section{Experimental} 
The studied single crystal of Bi2201 has a mass of 7 mg and approximate dimensions of $0.35 \times 0.3 \times 0.01cm^3$. The single crystal shows a fully developed superconducting transition at $T_c \approx 7$~K with $\Delta T_{c}\simeq 1.5$ K. The crystal was grown by a travelling-solvent-floating-zone method described in Ref.~\cite{luo}. The magnetization measurements were carried out by using a Cryogenics magnetometer system based on a superconducting quantum interference device, built with a 60 kOe superconducting magnet with a low-field feature. We only obtained measurements for the direction $H\|c$-axis. All data were obtained after  a desired magnetic field was applied without overshooting and magnetization data were collected by heating the sample with fixed $\delta T \leq 0.1$ K increments from 3 K up to temperatures well above $T_c$ producing a isofield zero-field-cooling (ZFC) $M(T)$ curve. After that the sample was again cooled to 3 K, in the applied magnetic field, and another set of data was collected while heating the sample from 3 K to above $T_c$, producing a isofield field-cooling (FC) $M(T)$ curve.  This procedure allows for determination of the reversible regime of each isofield $M~vs.T$ curve. Few isothermic $M(H)$ curves were also obtained which exhibit a pronounced fish-tail (second magnetization peak). 
As mentioned, the suppression of superconductivity observed with a 30 kOe field allowed to obtain the precise form of the normal state magnetisation, $M_{back}$= $a+bT+c/T$. The background magnetisation was then corrected for each $M(T)$ curve after selecting and fitting a wide region above $T_c$ to the form $a+bT+c/T$, where $a<0$, $b$ is small, positive for low fields, decreases as field increases changing signal for higher fields, and $c>0$.
%%%%%%%%%%%%%%%%%%%%%%%%%%%%
\section{Results and discussion}
Figure 1 shows all measured ZFC $M(T)$ curves obtained after the proper background correction. The arrows in Fig. 1 show the positions at which the ZFC and FC curves separate (producing a kind of hump) as $T$ is lowered below the irreversibility temperature, $T_{irr}$, which is only observed for fields $H\leq 1$ kOe. All curves obtained for fields above 1 kOe show only reversible regime above 3 K. The inset of Fig. 1 shows selected $M(T)$ curves as measured, evidencing the behavior of the $M(T)$ curve for $H=30$ kOe (for which no superconductivity signal is detected) and as measured for $H=20$ kOe, both exhibiting the same normal state behavior which are well fitted to the form $a+bT+c/T$. This inset also shows the resulting magnetisation curve for $H=20$ kOe obtained after subtraction of the normal state magnetisation extrapolated to the lower temperature region. For this particular curve ($H=20$ kOe), the normal state region was obtained by fitting the data above $T$ = 10 K, which produced the best fit. We mention that there is no visible changes in the resulting magnetisation after using data above 12 K in the fitting. For all curves we used the same criterium, by selecting the normal state region that produced the best fit. It should be mentioned that a similar normal state background with a small linear temperature dependence term was previously observed in Bi2201 nanocristaline-phase \cite{jin}.  As shown in the curves of Fig.~1, the resulting isofield $M(T)$ curves show a large reversible region displaying a linear region with $T$, followed by a pronounced rounding effect as magnetization approaches zero, which is reminiscent of superconducting fluctuations in the vicinity of $T_c(H)$ in high-$T_c$ superconductors.\cite{rosenstein,klemm,kes,tesa} It is also possible to see in Fig. 1 the existence of a crossing point defined by curves going from 500 Oe up to 15 kOe. 
\begin{figure}[t]
\includegraphics[width=\linewidth]{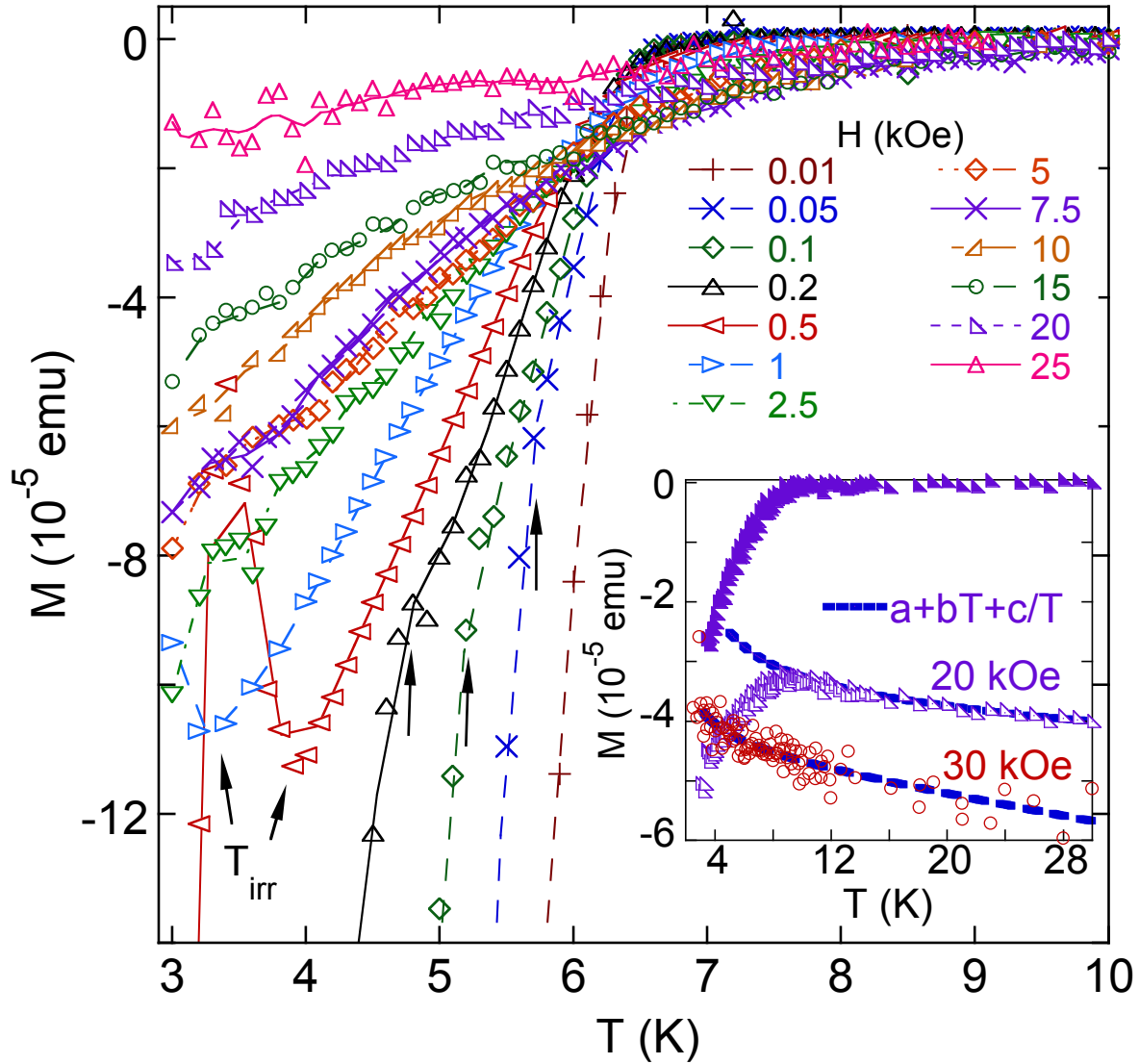}
\caption{Isofield ZFC $M~vs.~T$ curves measured with $H\|c$-axis. The inset shows selected $M~vs.~T$ curves (open symbols) prior background correction (dashed lines). As an example, closed symbols correspond to 20 kOe data after the background correction.}
 \label{fig1}
\end{figure}

Figure 2a shows details of the reversible region of selected $M(T)$ curves, where in this figure, it is plotted both, the ZFC and FC data. It is possible to see in Fig. 2a that the extrapolation of the linear region of each curve to $M=0$, which method\cite{abrikosov} is commonly used to estimate the mean field temperature transition, $T_c(H)$, produces inconsistent values of $T_c(H)$, which values even increase with field. 
Deviations from the linear behavior of $M(T)$ curves near $T_c(H)$ are examined by studying the asymptotic behavior\cite{s1,s2,s3} of $\sqrt {-M}$$\propto (T-T_a)^m$, where $m$ is an exponent (which mean field value is $1/2$ \cite{degennes}) and  $T_a(H)$ an apparent temperature transition. The approach is based on the fact that the magnetisation in the Abrikosov approximation, near $T_c(H)$, is given by 
$M=-\frac{e \hbar}{mc}|\psi_s|^2$ where $\psi_s$ is the superconducting order parameter.\cite{degennes} It follows that near $T_c(H)$, $\sqrt{-M} \propto [T_c(H)-T]^m$ where $m=1/2$ for  $s$-wave BCS superconductors \cite{degennes}. The fitting is performed by selecting a reversible region of data below the inflection point of each $\sqrt {-M}~vs.~T$ curve. The inflection point occur near a change in the concavity of the curve, above which amplitude fluctuations are dominant.  Results of this analysis are shown in Fig. 2b for selected $M(T)$ curves. Resulting values of $T_a$ decrease as field increases, but, as will be seen, lies above $T_c(H)$. Resulting values of the exponent $m$ deviates considerably from the expected mean field value, $1/2$, which suggests the existence of phase fluctuations \cite{kwon} (as discussed below), which extends phase correlations above $T_c(H)$ up to $T_a(H)$. The resulting values of $T_a(H)$ will be plotted below in a phase diagram of the studied sample. It is worth to mention that deviations from the mean field exponent value, 1/2, are expected for order parameters with $d$-wave symmetry (as is the case for cuprates), for which phase and amplitude fluctuations are predicted to have distinct contributions at the node and anti-node \cite{kwon}.  It is shown in Ref.~\cite{kwon} that in that case (the existence of a node in the order parameter) the effect of phase fluctuations is to reduce the density of states changing the shape of the gap near $T_c(H)$, and consequently the value of the exponent $m$. Within this scenarios, $T_a(H)$ 
represents the onset of phase correlations which may occur above $T_c(H)$. It is important to mention that the possible existence of a pseudo-gap phase above $T_c$ in Bi2201 would make one to expect values of $T_a(H)$ lying above $T_c$ \cite{beck}, as for instance observed in deoxygenated YBa$_2$Cu$_3$O$_{7-\delta}$ and Bi2212 (see Refs.~\cite{s1} and \cite{s2}), which is not the case here since the $T_a(H)$ line lies below $T_c$. This fact does not rule out the existence of a pseudo-gap above $T_c$ in the studied system, but with no phase correlations. It should mentioned that, due to the similarities between Bi2201 and Bi2212, the superconducting fluctuations observed here, could be explained in terms of uncorrelated pairs existing above $T_c$ due to charge inhomogeneity, as observed in Bi2212 within the pseudo-gap region \cite{gomes1,gomes2}. The inset of Fig. 2a shows a selected $M(H)$ curve obtained at 3 K, where it is possible to identify a Meissner region up to 3 Oe, and the field $H_{on}$ above which a pronounced second magnetisation peak sets in. The Meissner region allows to obtain $M$ in Gauss units.
\begin{figure}[t]
\includegraphics[width=\linewidth]{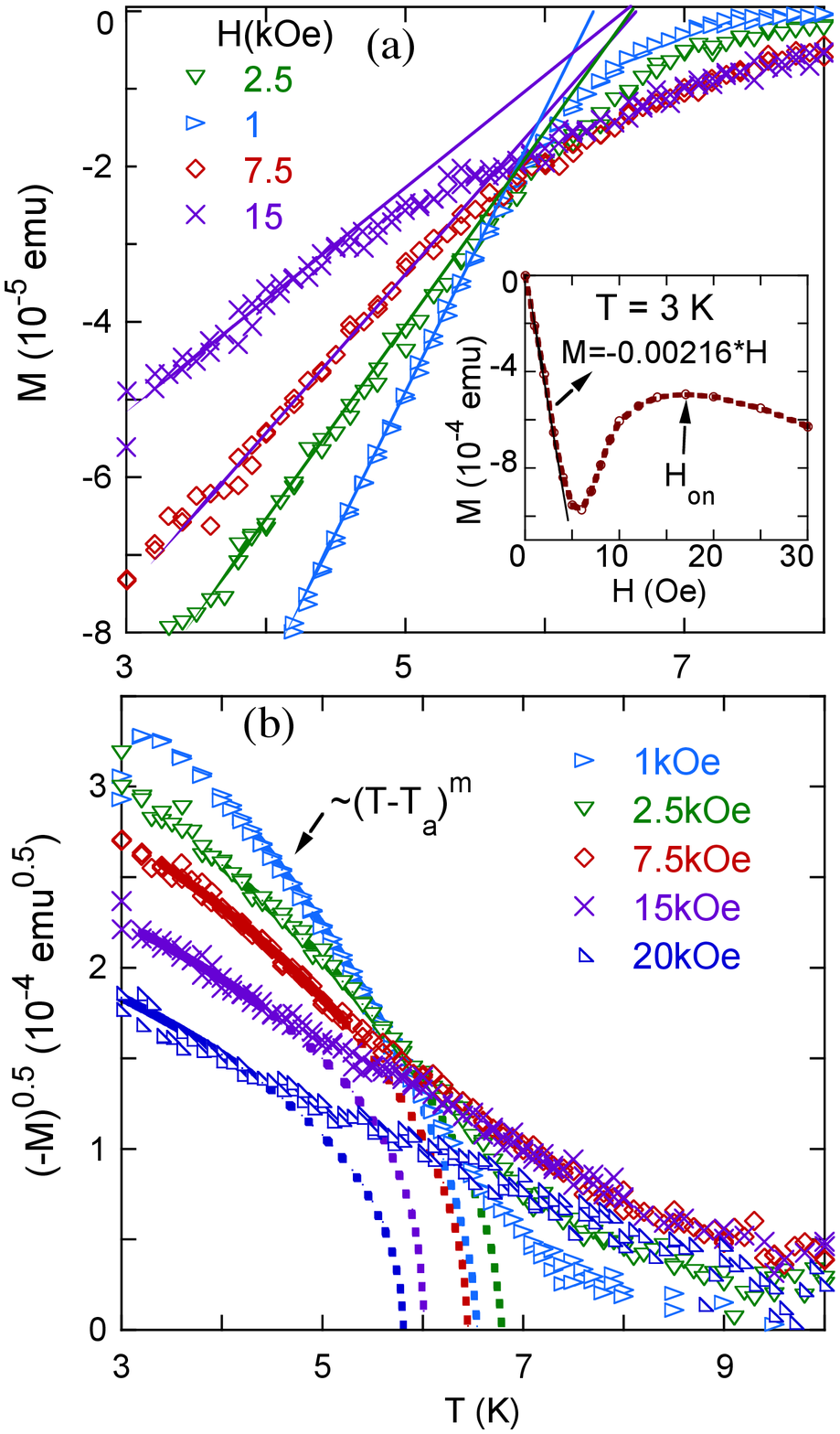}
\caption{a) Selected ZFC and FC isofield $M~vs.~T$ curves evidencing the linear extrapolation of $M$ until $M=0$. The inset show a low field plot of a $M(H)$ curve at 3 K. b) Selected isofield curves are plotted as $(-M)^{0.5}~vs.~T$. Solid lines show fittings of selected regions to the form $(T-T_a)^m$, and dotted lines show extrapolation of these fitting to $M~=~0$. Values of the exponent $m$ change from 0.40 for $H=20$~kOe to 0.57 for $H=1$ kOe.}
 \label{fig2}
\end{figure}
 
Figure 3 shows the analysis of the large temperature region above $T_a(H)$ which corresponds to amplitude fluctuations without phase correlations. The magnetization in this region is interpreted in terms of the Gaussian GL theory, which is extended to the case of high reduced-temperatures, $\varepsilon\equiv\ln(T/T_c)$, and high reduced-magnetic-fields $h\equiv H/H_{c2}'T_c$ [here $-H_{c2}'$ is the $H_{c2}(T)$ slope at $T_c$], by introducing a cut-off in the fluctuation spectrum to eliminate the contribution of the high-energy fluctuation modes.\cite{EPLVidal} According to this model, the fluctuation magnetic moment of single-layer two-dimensional superconductors is given by\cite{PC_Carlos}
\begin{eqnarray}
M=-\frac{k_BTV}{\phi_0s}\left[-\frac{c}{2h}\psi\left(\frac{h+c}{2h}\right)-\ln\Gamma\left(\frac{h+\varepsilon}{2h}\right)\right.\nonumber \\
+\left.\ln\Gamma\left(\frac{h+c}{2h}\right)+\frac{\varepsilon}{2h}\psi\left(\frac{h+\varepsilon}{2h}\right)+\frac{c-\varepsilon}{2h}\right].
\label{Prange2D}
\end{eqnarray}
Here $\Gamma$ and $\psi$ are the gamma and digamma functions, $k_B$ the Boltzmann constant, $\phi_0$ the flux quantum, $V\approx1$~mm$^3$ the sample volume, and $c\approx 0.5$ the total-energy cut-off constant.\cite{EPLVidal} This expression is applicable up to $\varepsilon=c$ (which corresponds to the reduced temperature at which fluctuation effects vanish), and up to $h\approx c/2\approx0.3$ \cite{PC_Carlos}. In the low magnetic-field limit ($h\ll\varepsilon$) Eq.~(\ref{Prange2D}) may be approximated by
\begin{equation}
M=-\frac{k_BTV}{6\phi_0s}h\left(\frac{1}{\varepsilon}-\frac{1}{c}\right),
\label{Schmidt2D}
\end{equation}
which is linear in $h$. In turn, in absence of cutoff (i.e., for $c\to\infty$) it reduces to the well known Schmidt-like expression for 2D materials, proportional to $\varepsilon^{-1}$.

\begin{figure}[t]
\includegraphics[width=\linewidth]{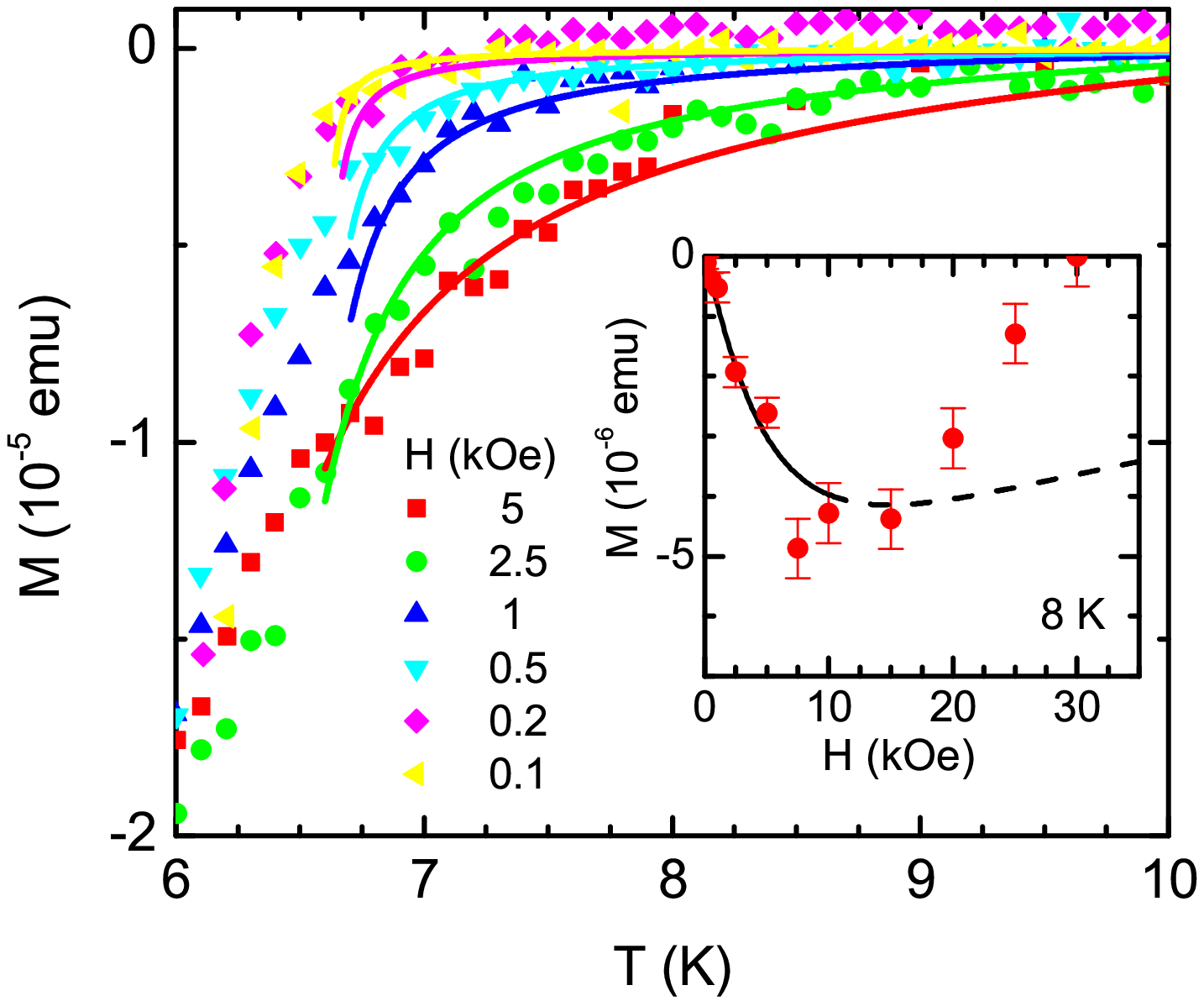}
\caption{Temperature dependence of the fluctuation-induced magnetic moment above $T_c$ for fields up to 30 kOe. The lines are the best fit of Eq.~(\ref{Prange2D}) with $T_c$ and $H_{c2}'$ as free parameters. Inset: example (corresponding to 8~K) of the magnetic field dependence of $M$. The line corresponds to Eq.~(\ref{Prange2D}) evaluated with the same parameters as in the main figure (it is printed as dashed beyond the applicability limit, $h\approx0.3$). It is worth noting that fluctuation effects vanish at $\sim30$~kOe, a value consistent with the upper critical field extrapolated to 0~K, as obtained from the resulting $T_c$ and $H_{c2}'$ values. }
 \label{fig1}
\end{figure}

The lines in Fig.~3 are the best fit of Eq.~(\ref{Prange2D}) to the $M(T)$ data for fields up to 5 kOe, well within the applicability range of Eq.~(\ref{Prange2D}) (see below). All curves were obtained by using $T_c$ and $H_{c2}'$ as free parameters. The fit is reasonably good in all the accessible temperature range above $T_c$, and leads to $T_c=6.6$~K and $H_{c2}'=4.5$~kOe/K. In the inset it is presented an example, corresponding to $T=8$~K, of the field dependence of the fluctuation magnetic moment. The line corresponds to Eq.~(\ref{Prange2D}) evaluated with the same parameters as in the main figure. It is interesting that, after the initial (roughly linear) increase of $|M|$ and the subsequent rounding associated to finite field effects, it begins to decrease and ends up vanishing at $H\approx30$~kOe. This field is consistent with the $H_{c2}$ value extrapolated to $T=0$~K, as may be estimated from the above $T_c$ and $H_{c2}'$ values, and from the analysis in the critical region (see below).

The existence of a line in the $H-T$ phase diagram at which fluctuation effects vanish has been shown few years ago in the magnetization of low-$T_c$ alloys \cite{breakdown,rukier}. In agreement with the present results, such a line was found to tend to $H\approx H_{c2}(0)$ (i.e., $h\approx1$) at low temperatures and to $T\approx 1.7T_c$ (i.e., $\varepsilon\approx0.5$) at low field amplitudes \cite{breakdown}. It was suggested that the vanishing of fluctuation effects at these high $h$ or $\varepsilon$ values is related to the shrinkage of the superconducting coherence length to its minimum possible length, the Cooper pairs size $\xi_0$, and that should be accounted for by the introduction of a total-energy cutoff in the fluctuations' spectrum \cite{breakdown}. 
In presence of large $h$ values such an introduction presents some difficulties \cite{breakdown,approximation} and, in fact, Eq.~(\ref{Prange2D}) is not applicable above $h\approx0.3$. However, very simple arguments \cite{breakdown} indicate that fluctuation effects should vanish at $h\approx1.1$, in good agreement with the present observations.

\begin{figure}[t]
\includegraphics[width=\linewidth]{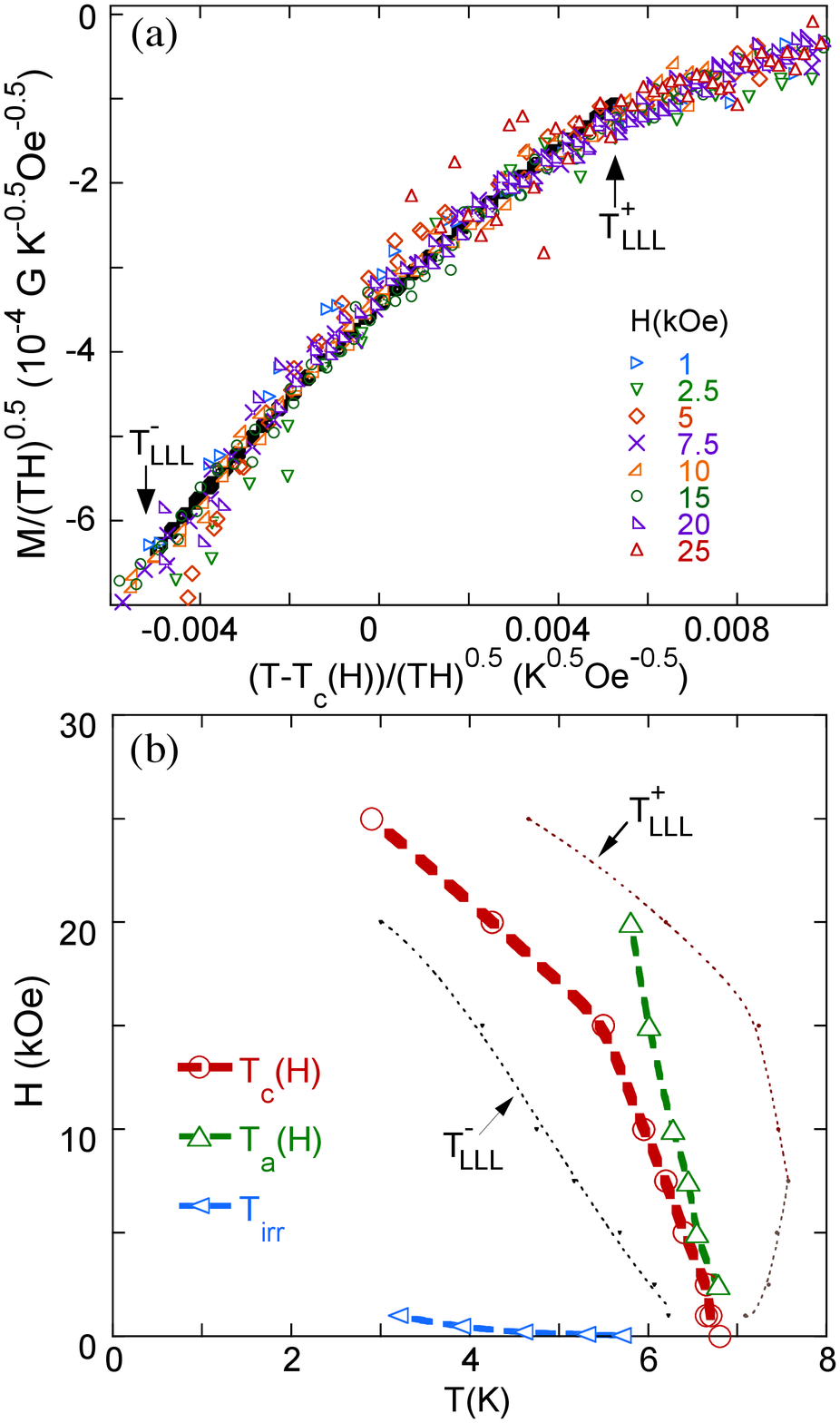}
\caption{a) Reversible isofield $M~vs.~T$ curves for $H\geq1$ kOe are plotted following the $2D$-LLL scaling law for magnetization. The dashed line represents a fitting of the selected region to the Eq. 7 of Ref. \cite{tesa}. b) The resulting phase diagram, where the lines (dashed and dotted) are only a guide to the eyes.}
 \label{fig4}
\end{figure}
 
An inspection of Fig. 2a  shows in better detail the existence of a crossing point of $M(T)$ curves occurring at $T^*\approx5.9$~K. Crossing points appearing in plots of many isofield $M(T)$ curves have been observed in  other high-$T_c$ systems\cite{rosenstein} and interpreted in terms of vortex fluctuations treated in a Ginzburg-Landau, GL, theory under a lowest-Landau-level, LLL, approximation which takes in account fluctuations correlations.\cite{ullah,tesa,tesa2,zacarias,rosenstein2} As discussed in Ref. \cite{tesa} vortex fluctuations are expected for low magnetic fields within the London region, while, for high fields amplitude fluctuations dominate, explaining why the curves with 20 and 25 kOe do not cross the same point (see Fig. 1). The LLL critical theories under the GL formalism, predict universal expressions for magnetization and other thermodynamic quantities, depending on the dimensionality, $D$, of the system. While the universal expressions allows one to fit experimental data and obtain values of intrinsic parameters, the associated scaling laws allow to obtain values of the mean field temperature transition of the scaled curves.\cite{rosenstein} Following the general scaling law for magnetisation curves, plots of $M/(TH)^{(D-1)/D}$ vs. $(T-T_c(H))/(TH)^{(D-1)/D}$ where $D$ is the dimensionality and $T_c(H)$ an adjusted parameter, should collapse in to a universal curve. The existence of only one crossing point in Fig. 1 (better shown in Fig. 2), suggests absence of dimensional crossover induced by magnetic field, as observed in deoxygenated YBa$_2$Cu$_3$O$_{7-\delta}$.\cite{rosenstein2,jltp} The results of Fig. 3 above, as well the similarity between $M(T)$ curves of Bi2201 and Bi2212\cite{kes} evidences the $2D$ character of the studied system.

We finally show in Fig. 4a the results of the $2D$-LLL-scaling applied to the reversible region of the curves of Fig. 1 with $H\geq1$~kOe. Figure 4a is obtained by only adjusting a value of $T_c(H)$ for each curve in a way that all curves collapse together. Considering the large amount of $M(T)$ curves, the wide range of magnetic fields values (1 - 25 kOe), and the separation of the curves as shown in Fig. 1, the collapse shown in Fig. 4a is impressive. We mention that we also tried to apply the $3D$ version of the scaling law for magnetisation to the curves of Fig. 1, which failed. The values of $T_c(H)$ obtained from the $2D$-scaling are used to plot a phase diagram of the system, which is shown in Fig. 4b, producing $dH_{c2}/dT=-12$ kOe/K, and a value of $H_{c2}(0)\approx30$ kOe. It is important to point out that the values of $T_c(H)$ could only be obtained through the $2D$-scaling procedure. We also plot in the phase diagram, values of $T_a(H)$ and values of $T_{irr}$. It is important to note that the phase diagram do not show any upward curvature. On the other hand, an upward curvature is present in the plot of $H~vs.T_a(H)$, which also would show a  $dH_{c2}/dT$ higher than -12 kOe/K. As discussed above, $T_a(H)$ represents the onset of phase correlations, which value is larger than the corresponding $T_c(H)$ but smaller than $T_c$. It is also important to mention that the value of $H_{c2}(0)$ estimated from the WHH formula is about twice the value suggested by the phase diagram.  This fact may suggest that the WHH formula overestimate the values of $H_{c2}(0)$ for high-$T_c$ systems. The resulting phase diagram shown in Fig. 4a, produced quite different values than those found in Ref. \cite{vedeneev} for a sample with $T_c$ $\approx$ 7 K, where $dH_{c2}/dT$ values varies from $\approx$ -27 kOe/K  to -50 kOe/K and $H_{c2}(0)$ from $\approx$ 150 kOe to 240 kOe (depending on the criterium used to determine the onset of superconductivity). Also, the authors of Ref.\cite{vedeneev} found that a suppression of superconductivity for a sample with $T_c$ $\approx$ 7 K is reached with a 250 kOe field, while here no superconductivity was observed for a 30 kOe field. Our finding, roughly agrees with the suppression of superconductivity  observed with a 50 kOe field found in Ref. \cite{chen} for a Bi2201 sample with $T_c$ $\approx$ 7 K. Similarly large values of $H_{c2}(T)$ (as the values found in Ref. \cite{vedeneev}), were found in Ref. \cite{french} from an analysis of magnetoconductivity fluctuations above $T_c$ in Bi2201. It is worth mentioning that the agreement between our work and that of Ref. \cite{french} lies on the two-dimensional character of the fluctuations above $T_c$. 

We also fit a selected region of data in Fig. 4a (see the dashed line) by the correspondent expression for magnetization developed for two dimensional systems  in Ref. \cite{tesa}. We used the equations 7 and 10 appearing in Ref. \cite{tesa}, where Eq. 10 is written as $(T_c-T^*)/T^*$= $b(dH_{c2}/dT)/(a'^2 \phi_0 s(U_0)^2)$ where $a'$ and $b$ are the original GL parameters, $\phi_0$ is the quantum flux, $s$ is the interlayer distance,  $U_0\approx 0.8$ and $T^*= 5.9$ K is the crossing point temperature. The fitting is conducted to data selected in a region of scaled-temperatures limited by $T^+$$_{LLL}$ and $T^-$$_{LLL}$ as indicated by the arrows in Fig.~4a, producing the following values for the fitting parameters: $s=11.9$  \AA\,, $T_c=6.2$ K and $dH_{c2}/dT=-9800$ Oe/K. The value of the interlayer distance $s$ is in good agreement with the listed value\cite{luo}, $s\approx12$ \AA\ and the value of $dH_{c2}/dT$ is in good agreement with the experimental  value obtained here, however, the value of $T_c$ is lower than the experimental value. The values of $T^+$$_{LLL}$ and $T^-$$_{LLL}$, which represents the temperature region where the $2D$-LLL equations well fitted the data, are obtained for each original $M(T)$ curve, and plotted in Fig.~4b. We mention that the temperature region where the rounding effect is observed in each curve virtually matches with the correspondent region delimited by the $T^+$$_{LLL}$ and $T^-$$_{LLL}$ lines in Fig. 4b, emphasizing the role of LLL-critical-fluctuations on the pronounced rounding effect of the $M(T)$ curves. 

\section{Conclusions}
In conclusion, we obtained the magnetic phase diagram of a Bi2201 sample with $T_c\approx7$~K for $H\|c$-axis, showing a $dH_{c2}/dT=-12$ kOe/K, a  $H_{c2}(0)\approx30$~kOe and no upward curvature. The value of  $H_{c2}(0)\approx60$~kOe, estimated from the WHH formula, is twice the value suggested by the phase diagram. Due to the pronounced rounding effect on $M(T)$ curves, values of $T_c(H)$ could only be obtained trough a $2D$-LLL-critical scaling analysis of many isofield magnetization curves. An analysis considering deviations from the linear behavior of $M$ with temperature, interpreted as due to phase fluctuations of the order parameter, produced values of an apparent temperature transition $T_a(H)$ which value is larger than the correspondent $T_c(H)$ but smaller than $T_c$. The temperature $T_a(H)$ is interpreted as the onset of phase correlations, and a plot of $H~vs.~T_a(H)$ show an upward curvature. The large amplitude fluctuation observed above $T_a(H)$ on $M(T)$ curves are explained in terms of a Gaussian-GL approach by the introduction of a total-energy cutoff in the fluctuation spectrum. We also observed that a 30 kOe field suppresses any signal of superconductivity above 3 K allowing to obtain the precise form of the normal state magnetisation used to correct all M(T) curves.
\section*{Acknowledgements}
SSS,ADA,CM and ES acknowledges support from CNPq and FAPERJ, and JM and JDD from the Spanish MICINN and ERDF \mbox{(FIS2010-19807)} and from the Xunta de Galicia (2010/XA043 and 10TMT206012PR).

\end{document}